\documentclass[twocolumn,showpacs]{revtex4}

\usepackage{graphicx}
\usepackage{dcolumn}
\usepackage{amsmath}
\begin{document}

\title{de~Haas-van~Alphen oscillations in quasi-two-dimensional underdoped cuprate superconductors in the canonical ensemble
}

\author{N.~Harrison$^1$ and S.~E.~Sebastian$^2$}

\affiliation{$^1$National High Magnetic Field Laboratory, LANL, Los Alamos, NM 87545\\
$^2$Cavendish Laboratory, Cambridge University, JJ Thomson Avenue, Cambridge CB3~OHE, U.K.}
\date{\today}

\begin{abstract}
We calculate the de~Haas-van~Alphen (dHvA) effect waveform using the canonical ensemble for different Fermi surface scenarios applicable to the underdoped cuprate superconductor YBa$_2$Cu$_3$O$_{6.5}$, in which quantum oscillations have recently been observed. The harmonic content of the dHvA waveform of the principal $F_\alpha\sim$~500~T frequency is consistent with the existence of a second thermodynamically dominant section of Fermi surface that acts primarily as a charge reservoir. Oscillations in the charge density to and from this reservoir are shown to potentially contribute to the observed large quantum oscillations in the Hall resistance. 
\end{abstract}
\maketitle

Layered electronic structures are known to be favorable for unconventional superconductivity in several families of compounds~\cite{schrieffer1,bauer1,takahashi1}. The quasi-two-dimensional Fermi surfaces of layered systems also yield magnetic quantum oscillations notably different from those in conventional three-dimensional metals. When the interlayer hopping energy of the quasiparticles $t_c$ becomes less than the cyclotron energy $\hbar eB/m^\ast$, increased Landau level degeneracy causes the quantum oscillations to depart from the Lifshitz-Kosevich (LK) theoretical form~\cite{lifshitz1}. The extent to which the chemical potential oscillates plays a central role in determining the de Haas-van Alphen (dHvA) waveform~\cite{shoenberg1,jauregui1,alexandrov1}. In effect, chemical potential oscillations become a valuable tool for probing the thermodynamic mass of additional Fermi surface sections~\cite{harrison1,jauregui1} through their effect as a `charge reservoir,' and have been shown to play an important role in layered organic metals~\cite{harrison1,harrison2,harrison3}. In this paper, we propose two-dimensionality to play an equally important role in determining the dHvA waveform and Hall resistance oscillations in underdoped cuprate superconductors~\cite{sebastian1,jaudet1,doiron1}. Importantly, the extent to which the chemical potential oscillates can help distinguish between Fermi surface scenarios consisting of an isolated pocket or of multiple sections.

\begin{figure}
\centering \includegraphics*[scale=0.31,angle=0]{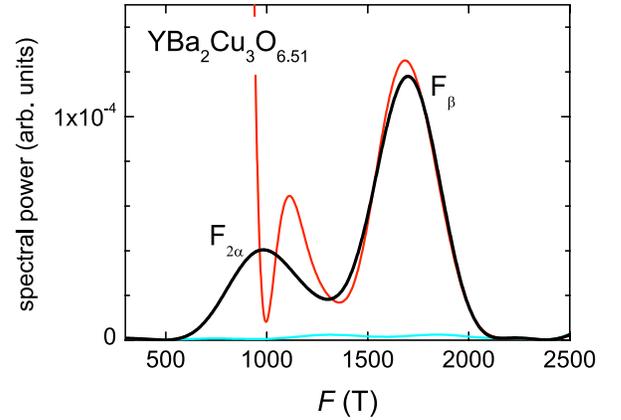}
\caption{dHvA oscillations suffer diffraction when Fourier transformed over a finite interval in magnetic field, with the side lobes sometimes interfering destructively with less prominent features. This situation can be remedied by subtracting fits to prominent features, such as the $F_\alpha\sim$~500~T frequency~\cite{doiron1}, prior to Fourier analysis. The black curve shows the result of such a procedure, yielding frequencies $F_{2\alpha}$ and $F_\beta$.  The red curve shows the same region of the Fourier transform prior to subtraction while the cyan curve shows the residual after subtracting fits to all three frequencies.
}
\label{powerspectrum}
\end{figure}

\begin{figure}
\centering \includegraphics*[scale=0.37,angle=0]{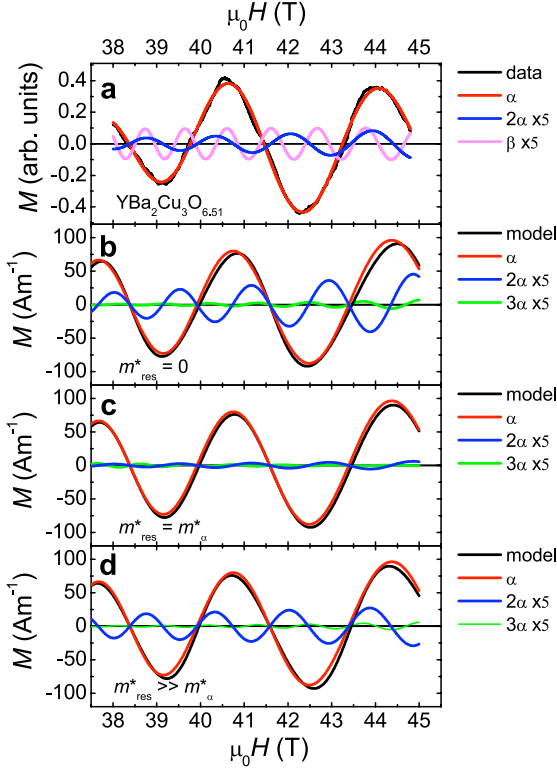}
\caption{Decomposition of the experimental data ({\bf a}) and numerical model simulations ({\bf b},{\bf c} and {\bf d}) into their constituent Fourier components. Note that the sign of the magnetization is opposite to that of the torque presented in Ref.~\cite{sebastian1}. The simulations are performed for a two-dimensional Fermi surface with background charge reservoir of different relative masses. The amplitudes of the harmonics and $F_\beta$ are increased $\times$~5 for clarity.
}
\label{waveform}
\end{figure}

The motivation for this study stems from the recent observation of a possible harmonic and second dHvA frequency in YBa$_2$Cu$_3$O$_{6.51}$~\cite{sebastian1}. These features are discernible in Fig.~\ref{powerspectrum} on performing a Fourier transform after subtracting a damped sinusoidal fit to the dominant $F_\alpha\sim$~500~T component~\cite{sebastian1}.  The lower frequency in Fig.~\ref{powerspectrum} is twice  $F_\alpha$, which is consistent with it being the harmonic $F_{2\alpha}$. Furthermore, on fitting the entire dHvA waveform to a sum $A=\sum_iA_i\sin(2\pi F_i/B+\phi_i)\exp(-\Gamma_i/B)$ over frequencies $i$ in Fig.~\ref{waveform}a, the $F_\alpha$ and $F_{2\alpha}$ oscillations can be seen cross through zero at similar values of the magnetic field$-$ thus exhibiting the key distinguishing feature of the dHvA effect in a two-dimensional metal~\cite{dHvAnote1}. 

To model the dHvA oscillations in YBa$_2$Cu$_3$O$_{6.51}$~\cite{sebastian1}, we therefore consider simplified quasi-two-dimensional dispersions $\varepsilon=\eta_i\hbar^2(k^2_x+k^2_y)/2m^\ast_i+2t_{c,i}(1-\cos[k_zc])$ for each Fermi surface section, where $\eta_i=+1$ for electrons and $-1$ for holes, $c$ is the interlayer spacing and $t_{c,i}\ll eB/m^\ast_i$. A large magnetic field along $z$ causes the density-of-states for each section $g_i[\varepsilon,B]=(\eta_ieB\Gamma_i/\pi^2\hbar c)\sum_\nu^\infty(\varepsilon^\prime_i[\nu]^2+\Gamma_i^2)^{-1}$ to become a series of Landau tubes $\nu$ where  $\varepsilon^\prime_i=\varepsilon-\varepsilon_{0,i}-\eta_i\hbar eB(\nu-1/2)/m_i^\ast\pm\Delta\varepsilon_i$~\cite{shoenberg1,harrison1}. Here, $\Delta\varepsilon_i$ represents a possible splitting of the Landau levels caused by Zeeman splitting~\cite{spinnote}, bilayer splitting or  $t_{c,i}$~\cite{splitting}. Scattering from random defects and impurities is assumed to broaden the Landau tubes into Lorentzians of width $2\Gamma_i=\hbar/\tau_i$~\cite{dingle1}.

Since the model treats values of $t_{c,i}\ll eB/m^\ast_i$, the entire Fermi surface contributes to the dHvA effect, requiring the magnetization $M=-\partial H_{\rm F}/\partial B|_{N,T}$ to be calculated numerically using the multiband canonical ensemble~\cite{harrison1,jauregui1,shoenberg1, alexandrov1}. The free energy is given by $H_{\rm F}=\sum_i\int^\infty_{-\infty}\varepsilon g_i[\varepsilon,B](1-\exp[\eta_i(\mu-\varepsilon)/k_{\rm B}T)])^{-1}{\rm d}\varepsilon$, requiring the chemical potential $\mu$ to be determined by fixing the total number of particles $N=\sum_i\int^\infty_{-\infty}g_i[\varepsilon,B](1-\exp[\eta_i(\mu-\varepsilon)/k_{\rm B}T)])^{-1}{\rm d}\varepsilon$.

We begin by considering a closed pocket ($\alpha$) in the presence of a background uniform density-of-states $g_{\rm res}=m^\ast_{\rm res}/\pi\hbar^2c$, which functions only as a charge reservoir (corresponding either to an open Fermi surface section or to a closed section with $\tau^{-1}\gg\tau^{-1}_\alpha$). Figures~\ref{waveform}b-d show calculations of the magnetization for background reservoirs of different effective masses, using $F_\alpha=$~499~T to match the phase of the oscillations measured in Fig.~\ref{waveform}a~\cite{phasenote},and  $m^\ast_\alpha=$~1.9~$m_{\rm e}$~\cite{doiron1}, $\Delta\varepsilon_\alpha=$~0 and  $\tau^{-1}_\alpha=$~2.5~$\times$~10$^{12}$~s$^{-1}$~\cite{scatteringnote}. The scenario depicted in Fig.~\ref{waveform}b corresponds to that of an isolated pocket (or two or more symmetry related pockets) with no charge reservoir present (i.e. $m^\ast_{\rm res}=$~0,see Refs.~\cite{doiron1,trugman1,damascelli1,leboeuf1}). However, on comparing Figs.~\ref{waveform}a and b, the amplitude of the $F_{2\alpha}$ frequency is this scenario can be seen to have the opposite sign of that observed experimentally~\cite{torqueinteraction}. In contrast, an additional charge reservoir of thermodynamic mass $m^\ast_{\rm res}\gtrsim m^\ast_\alpha$ (e.g. Fig.~\ref{waveform}d) can lead to simulated $F_{2\alpha}$ oscillations with the same sign as the experimental data.

\begin{figure}
\centering \includegraphics*[scale=0.31,angle=0]{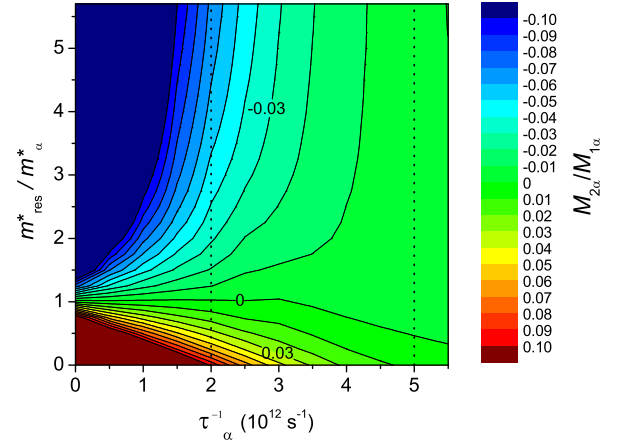}
\caption{Colored contour plot of the harmonic ratio $M_{2\alpha}/M_\alpha$ in the simulations versus $m^\ast_{\rm res}/m^\ast_\alpha$ and $\tau^{-1}_\alpha$ for the same average field range (41~T) as in Fig.~\ref{waveform}. The $M_{2\alpha}/M_\alpha=$~-0.03 contour corresponds most closely to that representing the data in Fig.~\ref{waveform}a while the dotted lines represent the likely upper and lower bounds of the experimental scattering rate (see text).
}
\label{contour}
\end{figure}

Figure~\ref{contour} shows the harmonic ratio $M_{2\alpha}/M_\alpha$ for the same average field interval as in Fig.~\ref{waveform} plotted for a variety of $\tau^{-1}_\alpha$ and $m^\ast_{\rm res}/m^\ast_\alpha$ values. The uncertainty in the contribution to the effective scattering rate from the superconducting pairing potential within the vortex state~\cite{corcoran1} requires us to consider a range of $\tau^{-1}_\alpha$ values. A precise estimate would require measurements at magnetic fields  $\mu_0H\gg\mu_0H_{\rm c2}\sim$~60~T~\cite{sebastian1}. Since the contribution to $\tau^{-1}$ from the vortex state is magnetic field-dependent~\cite{corcoran1}, estimates made from the gradient of a simple Dingle plot below $H_{\rm c2}$ typically over estimate the total extent of the scattering~\cite{sebastian1}. The value $\tau^{-1}_\alpha\approx$~5~$\times$~10$^{12}$~s$^{-1}$ obtained from such an analysis~\cite{dingle1,jaudet1,sebastian1} therefore can only be regarded as an upper bound (rightmost dotted line in Fig.~\ref{contour}). A lower bound estimate can be obtained from the extent to which the oscillations are damped between the end limits of the interval in magnetic field over which experiments are performed. From the oscillations of Jaudet {\it et al.}~\cite{jaudet1} measured between $\sim$~30 and 60~T, we obtain a lower bound of $\tau^{-1}_\alpha\approx$~2~$\times$~10$^{12}$~s$^{-1}$ (leftmost dotted line in Fig.~\ref{contour}). The region bounded by the dotted lines and by the uncertainty in the experimental harmonic ratio of $M_{2\alpha}/M_\alpha=$~-0.03~$\pm$~0.01 therefore provides us with the permissible range of parameters for modeling the dHvA effect, roughly corresponding to $\tau^{-1}_\alpha\lesssim$~3.5~$\times$~10$^{12}$~s$^{-1}$ and $m^\ast_{\rm res}/m^\ast_\alpha\gtrsim$~1.5.

Based on the above comparisons, rather than existing in isolation, the $\alpha$ pocket appears to coexist with a charge reservoir of thermodynamically greater mass. Open (or quasi-one-dimensional) Fermi surface sections originating from the Cu-O chains or from Fermi surface reconstruction by a collinear density-wave ordering with ${\bf Q}=(\pi,(1\pm 2\delta)\pi)$~\cite{millis1} provide potential candidates for a uniform reservoir of the type necessary to generate a waveform like that in Fig.~\ref{waveform}a. This would then require the $F_\beta$ oscillations to be the product of magnetic breakdown tunneling~\cite{sebastian1}.

\begin{figure}
\centering \includegraphics*[scale=0.4,angle=0]{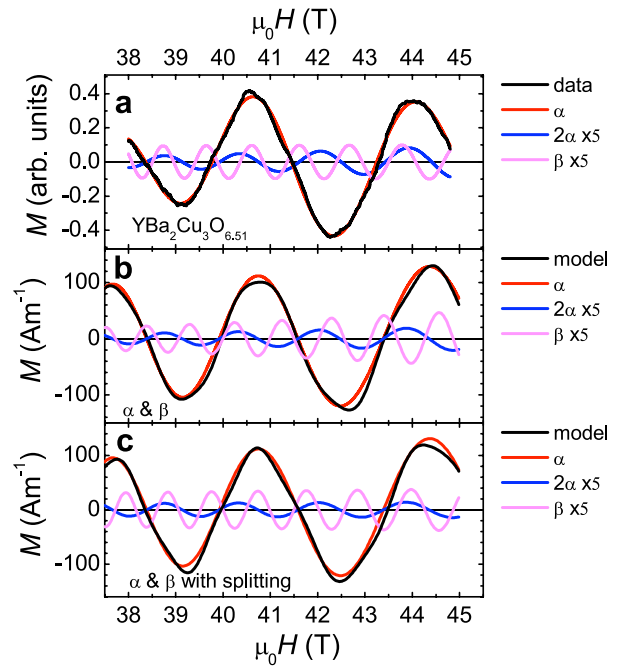}
\caption{Data re-plotted from from Fig.~\ref{waveform}a ({\bf a}) together with model calculations ({\bf b} and {\bf c}) for a Fermi surface consisting of a small electron pocket ($\alpha$) and large hole pocket ($\beta$). {\bf b} corresponds to $\tau^{-1}_\alpha=$~2.1~$\times$~10$^{12}$~s$^{-1}$, $\tau^{-1}_\beta=$~3.1~$\times$~10$^{12}$~s$^{-1}$ and $\Delta\varepsilon_\beta$=0 while {\bf c} corresponds to $\tau^{-1}_\alpha=$~2.1~$\times$~10$^{12}$~s$^{-1}$, $\tau^{-1}_\beta=$~2.6~$\times$~10$^{12}$~s$^{-1}$ and $\Delta\varepsilon_\beta=$~0.84~meV. 
}
\label{spiral}
\end{figure}

An alternative possibility is that the reservoir is composed of closed Fermi surface pockets of significantly greater thermodynamic mass or scattering rate than the $\alpha$ pocket so as to appear as relatively weak features in the dHvA signal. With its heavier effective mass of $m^\ast_\beta\approx$~3.8~$m_{\rm e}$, the $F_\beta$ frequency in Fig.~\ref{waveform}a (reproduced in Fig.~\ref{spiral}a) could correspond to this reservoir. A simulation including both $\alpha$ electron and $\beta$ hole pockets in Fig.~\ref{spiral}b with similar frequencies and effective masses to those reported in Ref.~\cite{sebastian1} can produce absolute amplitudes in similar proportions to those in Fig.~\ref{spiral}a on setting $\tau^{-1}_\alpha=$~2.1~$\times$~10$^{12}$~s$^{-1}$ and $\tau^{-1}_\beta=$~3.1~$\times$~10$^{12}$~s$^{-1}$. However, agreement with the sign of the observed $\beta$ frequency oscillations and the magnetic field dependence of its amplitude requires an additional effect, potentially provided by a fixed splitting term $\Delta\varepsilon_\beta=$~0.84~meV we introduce into the simulation in  Fig.~\ref{spiral}c  (also using $\tau^{-1}_\beta=$~2.6~$\times$~10$^{12}$~s$^{-1}$). This splitting could correspond to effects such as an interlayer tunneling $t_{c,\beta}=$~0.42 or residual bilayer splitting~\cite{sebastian1}. Such a value for $\Delta\varepsilon_\beta$ would give rise to a beat with nodes at $\sim$~18 and 55~T$-$ presently outside the field range of the torque experiments. The reduction in the amplitude of the $F_\beta$ oscillations with increasing field in YBa$_2$Ba$_3$O$_{6.50}$~\cite{sebastian1} could be consistent with a node at higher magnetic fields.

Another proposed scenario, corresponding to reconstruction of the Fermi surface by ${\bf Q}=(\pi,\pi)$, consists of hole pockets situated at ${\bf k}=(\pi/2,\pi/2)$ of nearly double the $k$-space area and frequency of proposed electron $\alpha$ pockets situated at ${\bf k}=(\pi,0)$~\cite{chakravarty1}. Since there are two hole pockets for every electron pocket in this scenario (therefore contributing twice as much to the dHvA signal and thermodynamic mass), the effective mass or scattering rate of the hole pocket would need to be several times that of the $\alpha$ pocket in order for its amplitude to be so much weaker in the experiment (e.g. requiring $m^\ast\gtrsim$~6~$m_{\rm e}$ or $\tau^{-1}\gtrsim$~7~$\times$~10$^{12}$~s$^{-1}$). 

\begin{figure}
\centering \includegraphics*[scale=0.3,angle=0]{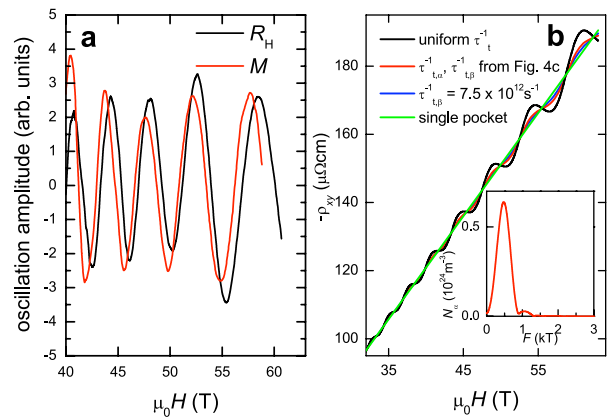}
\caption{({\bf a}) Oscillations in the Hall effect and magnetization from Refs.~\cite{jaudet1,doiron1}. The data have been rescaled by an exponential factor ${\rm e}^{-\gamma/B}$ to facilitate comparison. ({\bf b})The Hall resistivity due to $N_\alpha$ and $N_{\rm res}$ calculated using a two band model (assuming the $\alpha$ pocket to be electron-like) with non-oscillatory Drude conductivities determined by transport scattering rates as indicated. The black, red and blue curves have been renormalized by 42.2, 12.0 and 2.3 respectively. The inset shows a Fourier transform of $N_\alpha$ for 47~$<\mu_0H<$~62~T (roughly corresponding to the experimental range~\cite{doiron1}) for a reservoir consisting of a closed pocket ($\beta$) using the parameters as used in Fig.~\ref{spiral}c. Consistent with Hall resistance experiments, the $\beta$ frequency is not a prominent feature in $N_\alpha$.
}
\label{Hall}
\end{figure}

Quantum oscillations in the Hall resistance~\cite{doiron1} could provide further clues as to the nature of the charge reservoir. Doiron-Leyraud {\it et al.}~\cite{doiron1} find quantum oscillations that are a proportionately larger fraction of $R_{xy}$ than $R_{xx}$, suggesting a possible dominant oscillatory contribution to the Hall resistivity from oscillations in the carrier density $N_\alpha$ (and that of the reservoir) rather than $\sigma_{xx}$ or $\sigma_{yy}$~\cite{sigma}. Oscillations in the carrier density result from the oscillatory flow of charge back and forth between the pocket and the reservoir given by 
$N_\alpha-N_{\rm res}$, where $N_\alpha=\int^\infty_{-\infty}g_\alpha[\varepsilon,B](1-\exp[\eta_\alpha(\mu-\varepsilon)/k_{\rm B}T)])^{-1}{\rm d}\varepsilon$ and $N_{\rm res}=N_{{\rm res},0}-N_\alpha+N_{\alpha,0}$ and $N_{i,0}$ refers to the non oscillatory zero field values. Consistent with a dominant oscillatory carrier density contribution~\cite{reynolds1}, the phase of the oscillations in the experimental Hall resistance~\cite{doiron1} and magnetic torque~\cite{jaudet1} are aligned to within $\approx$~0.7~$\ll\pi/2$ in Fig.~\ref{Hall}a.

Figure~\ref{Hall}b shows the result of inserting the respective Hall coefficients $R_\alpha=1/eN_\alpha$ and $R_{\rm res}=1/eN_{\rm res}$ into a simple two-band expression for the Hall coefficient $R_{\rm H}=(R_\alpha\sigma_\alpha^2+R_\beta\sigma_{\rm res}^2)/(\sigma_\alpha+\sigma_{\rm res})^2$ on extending the simulations to higher magnetic fields. We neglect oscillations of $\sigma_{xx}=\sigma_{yy}$ by assuming simple Drude expressions $\sigma_{xx,i}=eN_{i,0}\tau_{i,{\rm t}}/m^\ast_i$, where $\tau^{-1}_{i,{\rm t}}\leq\tau^{-1}_i$ is the transport scattering rate. The oscillations are strongest (black curve) compared to the background on assuming a uniform transport scattering rate in the model including both $\alpha$ and $\beta$ pockets from Fig.~\ref{spiral}c, due to the very similar magnitudes of $R_\alpha\sigma_\alpha^2$ and $R_{\rm res}\sigma_{\rm res}^2$ in this case. They are also found to be prominent (red curve) on inserting $\tau^{-1}_{i,{\rm t}}=\tau^{-1}_i$ from Fig.~\ref{spiral}c. The oscillations become rather small (blue curve), however, if the reservoir has a poor conductivity by setting $\tau^{-1}_{\rm res,t}\geq$~7~$\times$~10$^{12}$ s$^{-1}$ or $m^\ast_{\rm res}\geq$~6~$m_{\rm e}$ as would be implied by the ${\bf Q}=(\pi,\pi)$ scenario discussed above, or if $R_{\rm res}=$~0 as for an ideal open Fermi surface. The oscillations vanish in the case of an isolated pocket with no reservoir (green curve). It should be noted, however, that the present simulations cannot reproduce the reported magnitude of the Hall resistivity~\cite{doiron1}. This would require a single isolated $\alpha$ electron pocket  with no Kramers degeneracy, the additional presence of a very small much higher mobility electron pocket or a possible vortex liquid contribution~\cite{chen1}.

In summary, by comparing the measured dHvA waveform of YBa$_2$Cu$_3$O$_{6.51}$ with canonical ensemble simulations, the sign of the harmonic of the dominant oscillatory $F_\alpha$ contribution suggests the coexistence of the $\alpha$ pocket with a thermodynamically dominant (i.e. heavier mass) charge reservoir. This finding is consistent with the the existence of multiple carrier sections~\cite{sebastian1,leboeuf1}. A simulation including both electron and hole closed pockets in the canonical ensemble appears to consistently explain the main aspects of both the dHvA waveform and the phase of the oscillatory Hall resistance on allowing for small adjustments in the transport scattering rates. In-situ measurements of magnetization and Hall resistivity should ultimately enable the relative contributions of the carrier density and diagonal conductivity to be better resolved.

This work is supported by US Department of Energy, the National Science Foundation and the
State of Florida. The authors acknowledge helpful comments from Mike Norman and Hae-Young Kee.


\begin{thebibliography}{99}
\bibitem{dHvAnote1} The sawtooth dHvA waveform in a two-dimensional metal can be expressed as the sum $M_{\rm 2D}=\sum_pA_p\sin(2\pi pF/B)$ (where $p$ refers to the harmonic index) in contrast to that in three-dimensional metals where $M_{\rm 3D}=\sum_pA_p\sin(2\pi pF/B\pm\pi(-1)^p/4)$~\cite{shoenberg1,harrison1}.
\bibitem{spinnote} Should spin remain a good quantum number within the low temperature high magnetic field phase of YBa$_2$Cu$_3$O$_{6.5}$, $m^\ast_\alpha\sim$~1.9~$m_{\rm e}$~\cite{doiron1} and $g\sim$~2 implies that Zeeman splitting results in approximate degeneracy of the $\nu\pm 1^{\rm th}$ spin-up and $\nu\mp 1^{\rm th}$ spin-down Landau levels~\cite{shoenberg1}, therefore having a negligible effect on the simulated oscillations upon its omission. Complete degeneracy may be established if the high magnetic field phase is a spin-density wave.
\bibitem{splitting} Since the $k_z$-dependence of the density-of-states has singularities at $\pm 2t_{c,i}$, setting $\Delta\varepsilon_i=t_{c,i}$ serves as a useful approximation for $\Gamma_i\gg t_{c,i}$.
\bibitem{phasenote} Small adjustments are made to $F_i$ to facilitate comparison with experimental data.
\bibitem{scatteringnote} The value of $\tau^{-1}_\alpha$ affects the magnitude of $F_{2\alpha}$ but not the phase.
\bibitem{torqueinteraction} Torque interaction~\cite{shoenberg1} effects produce a distorted waveform and $F_{2\alpha}$ harmonic with a similar phase as that in Fig.~\ref{waveform}b, opposite to that observed in the experiment.
\bibitem{sigma} Oscillations in $\sigma_{xx}$ and $\sigma_{yy}$, by contrast, are in quadrature with the magnetization~\cite{chakravarty1}.
\end{thebibliography}
\end{document}